\def\year{2021}\relax
\documentclass[letterpaper]{article} 
\usepackage{aaai21}  
\usepackage{times}  
\usepackage{helvet} 
\usepackage{courier}  
\usepackage[hyphens]{url}  
\usepackage{graphicx} 
\usepackage{graphicx}  
\usepackage{natbib}  
\usepackage{xcolor}
\usepackage{caption} 
\usepackage{amsmath,amssymb}
\usepackage{subcaption}
\DeclareMathOperator*{\maxi}{maximize}
\DeclareMathOperator*{\mini}{minimize}
\DeclareMathOperator*{\argmax}{arg\,max}
\DeclareMathOperator*{\argmin}{arg\,min}
\usepackage[ruled,linesnumbered]{algorithm2e}
\newcommand{\nonl}{\renewcommand{\nl}{\let\nl\oldnl}}
\makeatletter
\newcommand{\nosemic}{\renewcommand{\@endalgocfline}{\relax}}
\usepackage{multirow}

\title{Generative Inverse Deep Reinforcement Learning for Online Recommendation}
\author{
    Xiaocong Chen\textsuperscript{\rm 1}, Lina Yao\textsuperscript{\rm 1}, Aixin Sun\textsuperscript{\rm 2}, Xianzhi Wang\textsuperscript{\rm 3}, Xiwei Xu\textsuperscript{\rm 4}, Liming Zhu\textsuperscript{\rm 4} \\
}

\affiliations{
    \textsuperscript{\rm 1}School of Computer Science and Engineering, University of New South Wales, Australia \\
    \textsuperscript{\rm 2}School of Computer Science and Engineering, Nanyang Technological University, Singapore \\
    \textsuperscript{\rm 3}School of Computer Science, University of Technology Sydney, Australia\\
    \textsuperscript{\rm 4}Data61, CSIRO, Sydney, Australia\\

    \{xiaocong.chen, lina.yao\}@unsw.edu.au, axsun@ntu.edu.sg,xianzhi.wang@uts.edu.au, \{Xiwei.Xu, Liming.Zhu\}@data61.csiro.au
}

\begin{document}
\maketitle

\begin{abstract}
    Deep reinforcement learning enables an agent to capture user's interest through interactions with the environment dynamically. It has attracted great interest in the recommendation research.
    Deep reinforcement learning uses a reward function to learn user's interest and to control the learning process. However, most reward functions are manually designed; they are either unrealistic or imprecise to reflect the high variety, dimensionality, and non-linearity properties of the recommendation problem. That makes it difficult for the agent to learn an optimal policy to generate the most satisfactory recommendations. To address the above issue, we propose a novel generative inverse reinforcement learning approach, namely InvRec, which extracts the reward function from user's behaviors automatically, for online recommendation. We conduct experiments on an online platform, VirtualTB, and compare with several state-of-the-art methods to demonstrate the feasibility and effectiveness of our proposed approach. 
\end{abstract}

\section{Introduction}
Deep reinforcement learning (DRL) is promising for recommendation systems, given its ability to learn optimal strategies from interactions for generating recommendations that best fit users' dynamic preferences. DRL-based recommendation systems cover three categories: deep Q-learning based methods, policy gradient based methods, and hybrid methods.
Deep Q-learning aims to find the best step by maximizing a Q-value over all possible actions. \citet{zheng2018drn} first introduced DRL into recommendation systems for news recommendation; then, \citet{chen2018stabilizing} introduced a robust Q-learning to handle dynamic environments for online recommendation. However, Q-learning based methods suffer the \textit{agent stuck problem}, i.e., Q-learning requires the maximise operation over the action space, which becomes infeasible when the action space is extremely large.
Policy gradient based methods can mitigate the agent stuck problem~\cite{chen2019large}. Such methods use average reward as the guideline yet they may treat bad actions as good actions, making the algorithm hard to converge~\cite{pan2019policy}.
In comparison, hybrid methods combine the advantages of Q-learning and policy gradient. As a popular algorithm among hybrid methods, actor-critic network~\cite{konda2000actor} adopts policy gradient on an actor network and Q-learning on a critic network to achieve nash equilibrium on both networks. Until now, actor-critic network has been widely applied to DRL-based recommendation systems ~\cite{chen2019top,liu2020end,zhao2020leveraging,chen2020knowledge}.

\begin{figure*}[ht]
    \begin{subfigure}{0.49\linewidth}
        \includegraphics[width=\linewidth]{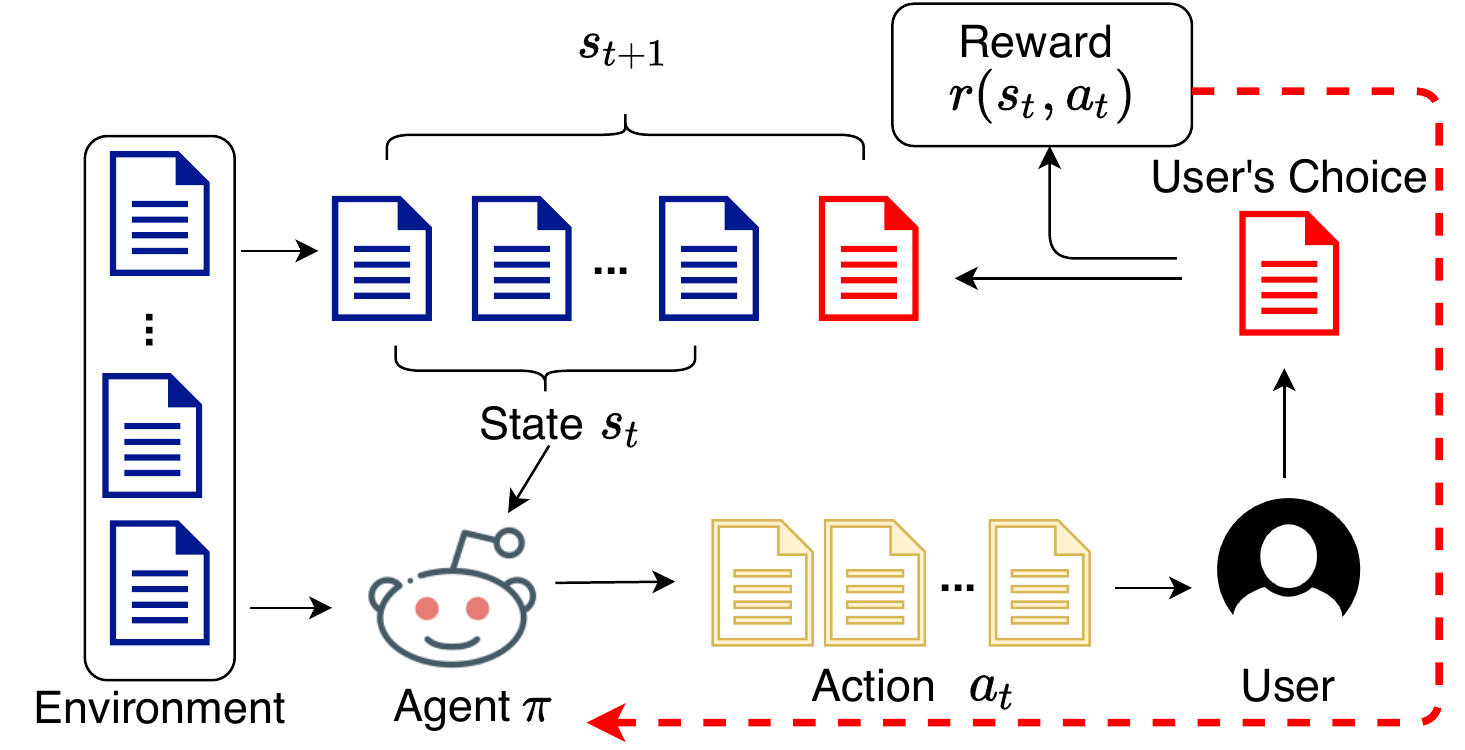}
        \caption{}
    \end{subfigure}
    \begin{subfigure}{0.49\linewidth}
        \includegraphics[width=\linewidth]{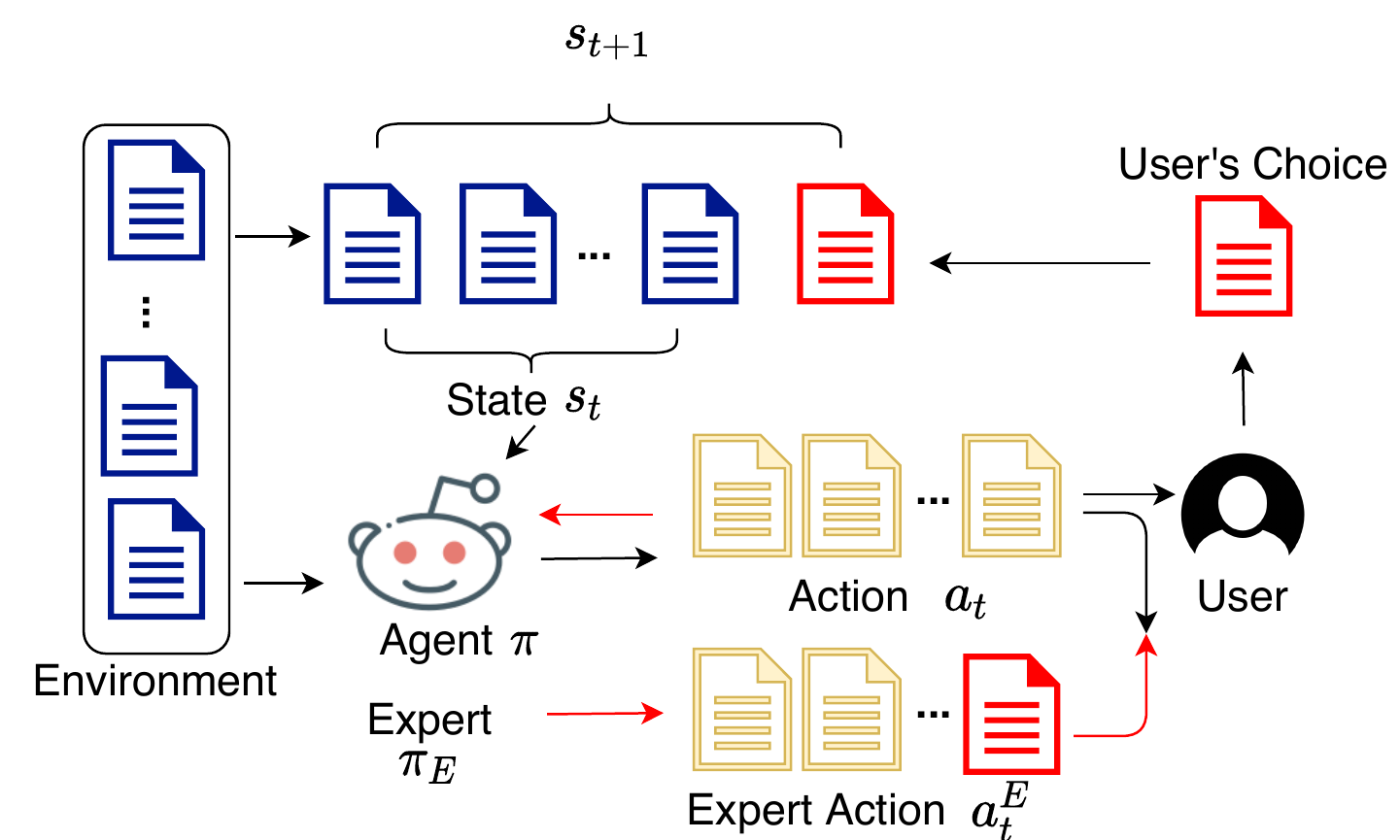}
        \caption{}
    \end{subfigure}
    \caption{General workflow of current reinforcement learning based recommendation system (a) which relies on reward function to guide the agent. The inverse reinforcement learning based approach (b) which does not require the reward function guiding. We use red color to represent how those two approaches update the policy.}
    \label{fig:work}
\end{figure*}

Despite differences, all existing DRL-based methods rely on well-designed context-dependent reward functions, as shown in the general workflow of a DRL-based recommendation systems in Fig~\ref{fig:work} (a).
However, in many cases, the reward function cannot be easily defined, due to dynamic environments and various factors that affect user's interest ~\cite{shang2019environment}.
Thus, the existing methods may
suffer limited generalization ability. Besides, they generally take into account both user's preference and user's actions (e.g., click-through, rating or implicit feedback) in the reward function, under the assumptions that the reward is determined by the current chosen item and user's action is unaffected by the recommended items. Such assumptions, however, no longer hold for online recommendation~\cite{shang2019environment}.
Another challenge is that the usage of reinforcement learning in finding the recommendation policy from scratch might be time-consuming~\cite{finn2016guided,levine2012continuous}. In a recommendation problem, the state space (i.e., all candidate items in which users might be interested) and action space (i.e., all the actions for candidate items) might be huge, traditional reinforcement learning based methods iterate all the possible combinations to figure out the best policy which is arduous.  

Targeting at the above challenges, we aim to enable the agent to infer a reward function from user's behaviors via inverse reinforcement learning and learn the recommendation policy directly from user behaviors in an efficient and adaptive way, as shown in Fig 1 (b).
To this end, we propose a generative inverse reinforcement learning approach for adaptively inferring an implicit reward function from user behaviors.
Specifically, the approach can measure the performance of the current recommendation policy and update the current policy with the expert policy in the discriminator, thus alleviating the need of defining reward function for online recommendation.
In particular, we transform inverse deep reinforcement learning (DRL) into a generator to augment a diverse set of state-action pairs.
Under this generative strategy, our method can achieve better generalization ability under complex online recommendation conditions. In a nutshell, we make the following contributions: 
\begin{itemize}
    \item We propose generative inverse reinforcement learning to automatically learn reward function for online recommendation.
    To the best of our knowledge, this is the first work to decouple the reward function and the agent for online recommendation.
    \item We design a novel actor-discriminator network module that takes a discriminator as the critic network and a novel actor-critic network as the actor network, to implement the proposed framework. The module is model-free and can be easily generalized to a variety of scenarios. 
    \item We conduct experiments on a virtual online platform, VirtualTB, to demonstrate the feasibility and effectiveness of the proposed approach. Our proposed method achieves a higher click-through-rate that several state-of-the-art methods. 
\end{itemize}

\section{Problem Formulation}

\noindent{\bf Online Recommendation}
Online recommendation differs from offline recommendation in dealing with real-time interactions between users and the recommendation system. The system needs to analyse user's behavior and updates the recommend policy dynamically. The objective is to find a solution that best reflects those interactions and apply it to the recommend policy.

\noindent{\bf Reinforcement Learning-based Recommendation}
Reinforcement Learning based recommendation systems learn from interactions through an Markov Decision Process (MDP). 
Given a recommendation problem consisting of a set of users $\mathcal{U} = \{u_0,u_1,\cdots u_n\}$, a set of items $\mathcal{I} = \{i_0,i_1,\cdots i_m\}$ and user's demographic information $\mathcal{D}=\{d_0,d_1,\cdots,d_n\}$, MDP can be represented as a tuple $(\mathcal{S},\mathcal{A},\mathcal{P},\mathcal{R},\gamma)$, where $\mathcal{S}$ denotes the state space, which is the combination of the subsets of $\mathcal{I}$; $\mathcal{A}$ is the action space, which represents agent's selection during recommendation based on the state space $\mathcal{S}$; $\mathcal{P}$ is the set of transition probabilities for state transfer based on the action received; $\mathcal{R}$ is a set of rewards received from users, which are used to evaluate the action taken by the recommendation system, with each reward being a binary value to indicate user's click; $\gamma$ is a discount factor $\gamma \in [0,1]$ for balancing the future reward and current reward.

Given a user $u_0$ and the state $s_0$ observed by the agent (or the recommendation system), which includes a subset of item set $\mathcal{I}$ and user's demographic information $d_0$, a typical recommendation iteration for user $u_0$ goes as follows.
First, the agent makes an action $a_0$ based on the recommend policy $\pi_0$ under the observed state $s_0$ and receives the corresponding reward $r_0$.
Then, the agent generates a new policy $\pi_1$ based on the received reward $r_0$ and determines the new state $s_1$ based on the probability distribution $p(s_{new}|s_0,a_0)\in\mathcal{P}$.
The cumulative reward after $k$ iterations is as follows:
\begin{align*}
    r_c = \sum_{k=0} \gamma^{k}r_k
\end{align*}

\noindent{\bf Inverse Reinforcement Learning-based Online Recommendation}
We propose inverse reinforcement learning without predefining a reward function for online recommendation.
We aim to optimize the current policy $\pi:\mathcal{S}\to\mathcal{A}$ to make recommendations that are most suitable for the user.
\begin{figure*}[ht]
    \centering
    \includegraphics[width=\linewidth]{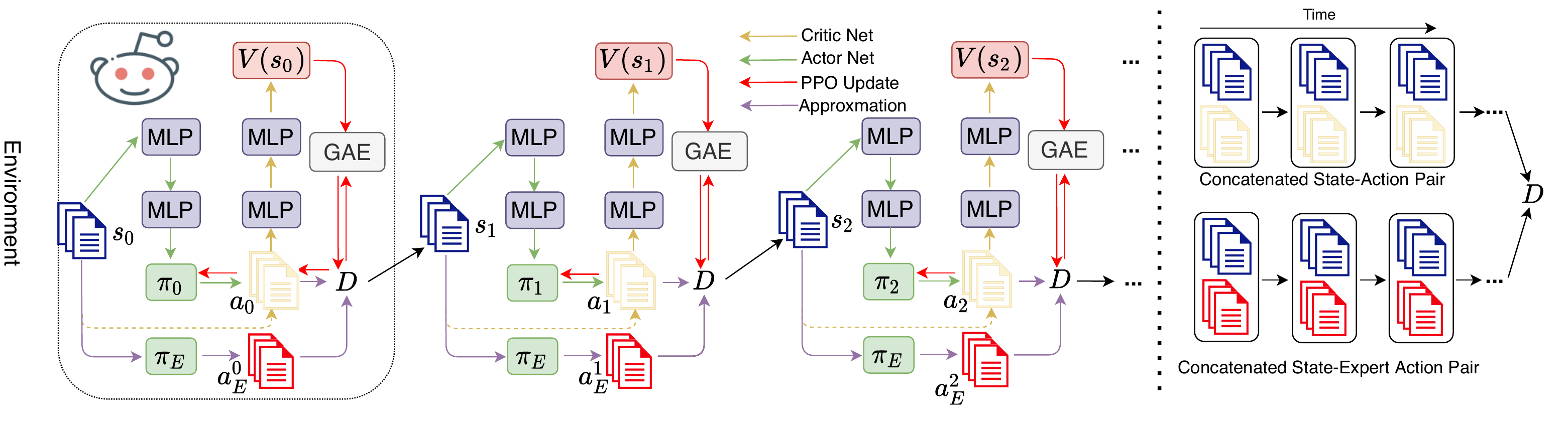}
    \caption{The proposed framework. The actor-critic network generates the policy $\pi$ based on current state $s_t$. The discriminator takes state-action pair $(s_t,a_t)$ and expert state-action pair $(s_t,a_E^t)$ as input. The goal of the actor-critic network is to generate $\pi$ which can let discriminator classify it as an expert policy. The discriminator aims to distinguish the policy that is generated by actor-critic network. The input of GAE is the Value from the critic network and the $c(s,a)$ from discriminator which details will be presented later. We circled one update episode which happened on agent.}
    \label{fig:structure}
\end{figure*}
%
Specifically, we model online recommendation as an MDP with a finite state set $\mathcal{S}$, a set of actions $\mathcal{A} = \{a_0, a_1, \cdots , a_t\}$, transition probabilities $\mathcal{P(\mathcal{S},\mathcal{A})}$, and a discount factor $\gamma$.
Suppose there exist an expert policy $\pi_E$ that can master any state $s\in\mathcal{S}$. The recommendation turns into the optimization problem of finding the policy $\pi$ that best approximates the expert policy $\pi_E$ across the cost function class $\mathcal{C}$, by the following objective function~\cite{abbeel2004apprenticeship}:
\begin{align}
    \mini_{\pi} \max_{c\in\mathcal{C}} \mathbb{E}_{\pi}[c(s,a)] - \mathbb{E}_{\pi_E}[c(s,a)]\label{eq1}
\end{align}
The cost function class $\mathcal{C}$ is restricted to convex sets defined by the linear combination of a few basis functions \{$f_1,f_2, \cdots, f_k$\}. Hence, the corresponding feature vector for the state-action pair $(s,a)$ can be represented as $f(s,a) = [f_1(s,a),f_2(s,a),\cdots,f_k(s,a)]$. The expectation $\mathbb{E}_{\pi}[c(s,a)]$ is defined as (on $\gamma$-discounted infinite horizon):
\begin{align}
    \mathbb{E}_{\pi}[c(s,a)] = \mathbb{E}[\sum_{t=0}^{\infty}\gamma^t c(s_t,a_t)]
\end{align}

\section{Methodology}
The overall structure of our proposed approach is illustrated in Fig~\ref{fig:structure}. It consists of three main components: policy approximation, policy generation, and discriminative actor-critic network.
\textit{Policy approximation} provides the theoretical approach to approximating the learned recommendation policy with the expert policy $\pi_E$;
\textit{policy generation} increases the diversity of the recommendation policy;
and \textit{discriminative actor-critic network} constitutes the main structure of our approach.
Besides the above components, we will present the optimization method and the corresponding training algorithm, which aims to limit the update step in optimizing the loss function to ensure that a new policy achieves better performance than the old one.

\subsection{Policy Approximation}
The policy approximation component aims to make the learned recommendation policy
$\pi$ and expert policy $\pi_E$ as similar as possible. To this end, we look for a cost function that delivers the best expert policy $\pi_E$ among all the policies on the latent space.
According to Eq.(\ref{eq1}), the cost function class $\mathcal{C}$ is convex sets, which have a linear format~\cite{abbeel2004apprenticeship} and a convex format~\cite{syed2008apprenticeship}, respectively:
\begin{align}
    & \mathcal{C}_{linear} = \{\sum_i w_if_i: \|w\|_2 \leq 1\} \\
    & \mathcal{C}_{convex} = \{\sum_i w_if_i: \sum_i w_i = 1, \forall i \text{ s.t. } w_i \geq 0\}
\end{align}
The corresponding objective functions are as follows:
\begin{align}
    & \max_{c\in \mathcal{C}_{linear}} \|\mathbb{E}_\pi[f(s,a)] - \mathbb{E}_{\pi_E}[f(s,a)]\|_2  \label{eq2}\\ 
    & \max_{j\in \{1,\cdots,k\}} \mathbb{E}_\pi[f_j(s,a)] - \mathbb{E}_{\pi_E}[f_j(s,a)] \label{eq3}
\end{align}

In particular, Eq.(\ref{eq2}) minimizes the $l_2$ distance between the state-action pairs, known as feature expectation matching~\cite{abbeel2004apprenticeship}.
Eq.(\ref{eq3}) minimizes the worst-case excess cost among the functions $f_k$~\cite{syed2008game}.
An issue with such methods is the ambiguity in Eq.(1) that many candidate policies $\pi$ can approximate the expert $\pi_E$ if we only compare the features~\cite{ziebart2008maximum}. We resolve the ambiguity by introducing the following $\gamma$-discounted causal entropy~\cite{bloem2014infinite} into Eq.(\ref{eq1}):
\begin{align}
     H(\pi) \triangleq  & \mathbb{E}_{\pi}[-\log\pi(a|s)] \nonumber\\ = & \mathbb{E}_{s_t,a_t\sim\pi}\bigg[-\sum_{t=0}^\infty\gamma^t\log\pi(a_t|s_t)\bigg] \label{eq7}
\end{align}
We thereby rewrite Eq.(\ref{eq1}) into
\begin{align}
    \mini_{\pi}-H(\pi) - \mathbb{E}_{\pi_E}[c(s,a)] + \max_{c\in\mathcal{C}}\mathbb{E}_{\pi}[c(s,a)] \label{eq8}
\end{align}
and define the reinforcement learning process according to~\cite{ziebart2008maximum}:
\begin{align}
    \argmin_{\pi \in \Pi} -H(\pi) + \mathbb{E}_{\pi}[c(s,a)]
\end{align}

Suppose $\Pi$ is the policy set. We define the loss function $c(s,a)$ such that the expert policy has a lower cost and other policies have higher cost.
Referring to Eq.(\ref{eq8}), we define the maximum causal entropy inverse reinforcement learning~\cite{ziebart2010modeling} as follows:
\begin{align}
    \maxi_{c\in\mathcal{C}} (\min_{\pi \in \Pi} -H(\pi) + \mathbb{E}_{\pi}[c(s,a)])  - \mathbb{E}_{\pi_E}[c(s,a)]  \label{eq10}
\end{align}

\subsection{Policy Generation}
We regard policy generation as the problem of matching two occupancy measures and solve it by training a Generative Adversarial Network (GAN)~\cite{goodfellow2014generative}.
The occupancy measure $\rho$ can be defined as:
\begin{align}
    \rho_{\pi}(s,a) = \pi(s|a)\sum_{t=0}^\infty \gamma^t P(s_t=s|\pi)
\end{align}
Since the generator aims to generate the policy as similar to the expert policy as possible, we use GAIL~\cite{ho2016generative} to bridge inverse reinforcement learning and GAN by making an analogy from the occupancy matching to distribution matching. Specifically, we introduce a GA regularizer to restrict the entropy function:
\begin{align}
    \psi_{GA}(c) = 
    \begin{cases} 
      \mathbb{E}_{\pi_E}[g(c(s,a))] & c < 0 \\
      \infty & \text{otherwise} \\
   \end{cases}
\end{align}
where $g(x)$ is defined as:
\begin{align}
    g(x) = 
    \begin{cases} 
      -x-\log(1-\exp(x)) & c < 0 \\
      \infty & \text{otherwise} \\
   \end{cases}
\end{align}

By introducing the GA regularizer, we can directly measure the difference between the policy and expert policy without needing to know the reward function.
We use the loss function from the discriminator as $c(s,a)$ in Eq.(\ref{eq10}).
We represent the negative log loss for the binary classification to distinguish the policy $\pi$ and $\pi_E$ via state-action pairs. The optimal of Eq.(\ref{eq14}) is equivalence to the Jensen-Shannon divergence~\cite{nguyen2009surrogate}:
\begin{align}
    \psi_{GA}(\rho_\pi - \rho_{\pi_E}) = \max_{D\in(0,1)^{\mathcal{S}\times\mathcal{A}}}\mathbb{E}_{\pi}[\log D(s,a)] \nonumber \\ + \mathbb{E}_{\pi_E}[\log (1-D(s,a))] \label{eq14}
\end{align}
\begin{align}
    D_{JS}(\rho_\pi,\rho_{\pi_E}) = D_{KL}(\rho_\pi\|(\rho_\pi+\rho_{\pi_E})/2) \nonumber + \\ D_{KL}(\rho_{\pi_E}\|(\rho_\pi+\rho_{\pi_E})/2)  \label{eq15}
\end{align}

Finally, We obtain the inverse reinforcement learning definition by substituting the GA regularizer into Eq.(\ref{eq8}):
\begin{align}
    \mini_{\pi}-\lambda H(\pi) + \underbrace{\psi_{GA}(\rho_\pi - \rho_{\pi_E})}_{D_{JS}(\rho_\pi,\rho_{\pi_E})} \label{eq16}
\end{align}
where $\lambda$ is a factor with $\lambda \geq 0$.
Note that Eq.(\ref{eq16}) has the same goal as the GAN, i.e., finding the squared metric between distributions.
More specifically, we have the following equivalence for Eq.(\ref{eq16}):
\begin{align}
    & \mini_{\pi}-\lambda H(\pi) + \psi_{GA}(\rho_\pi - \rho_{\pi_E})  \equiv \nonumber \min_{\pi}\max_{D} \mathcal{L_D}\\
    & \mathcal{L_D}=\mathbb{E}_{\pi}[\log D(s,a)] + \mathbb{E}_{\pi_E}[\log (1-D(s,a))] -\lambda H(\pi) \label{eq17}
\end{align}

\subsection{Discriminative Actor-Critic Network}
The discriminative actor-critic network aims to map online recommendation into an inverse reinforcement learning framework.

Specifically, we take \textit{advantage actor-critic network}, a variant of the actor-critic to constitute the main structure of our approach.
Within this network, the actor uses the policy gradient to update the policy, and the critic uses Q-learning to evaluate the policy and provides feedback~\cite{konda2000actor}.

Given user's profile at timestamp $t$ (i.e., the item list $\{i_0,\cdots,i_{t-2},i_{t-1}\}$) and the optional demographic information $\mathcal{D}$ (which is used to generate the state $s_t$), the environment embeds user's recent interest and user's features into the latent space via neural network~\cite{chen2020knowledge,liu2020end}.
Once the actor network gets the state $s_t$ from the environment, it feeds the state to a network with two fully-connected layers with ReLU as the activation function. The final layer outputs the target policy function $\pi$ parameterized by $\theta$, which will be updated together with discriminator $D$.
Then, the critic network takes the input from the actor network with current policy $\pi_{\theta_t}$, which can be used for sampling to get the trajectory $(s_t,a_t)$. We concatenate the state-action pair and feed it into two fully-connected layers with ReLU as the activation function. The output of the critic network is a value $V(s_t,a_t)\in {\rm I\!R}$, which will be used to calculate the advantage, which is a value used for optimization (to be discussed later).

As aforementioned, the discriminator $D$ is the key component of our approach. To build an end-to-end model and better approximate the expert policy $\pi_E$, we parameterized the policy as $\pi_\theta$ and clip the output of the discriminator so that $D:\mathcal{S}\times\mathcal{A} \to (0,1)$ with weight $w$. The loss function of $D$ is $\mathcal{L}_D$.
Besides, we use Adam~\cite{kingma2014adam} to optimize weight $w$ (the optimization for $\theta$ will be introduced later). Here, the discriminator $D$ can be treated as a local cost function to guide the policy update. 
Specifically, the policy will move toward expect-like regions (divided by $D$) in the latent space by minimizing the loss function $\mathcal{L}_D$, i.e., finding a point $(\pi,D)$ for Eq.(\ref{eq17}) such that the equation output is minimal.

\subsection{Policy Optimization}
We use the actor-critic network as a policy network to be trained jointly with the discriminator. Therefore, the actor-critic network needs to update the policy parameter $\theta$ based on the discriminator.
During this process, we aim to limit the agent's step size to ensure the new policy is better than the old one.
Specifically, we use trust region policy optimization (TRPO)~\cite{schulman2015trust} to update the policy parameter $\theta$ and formulate the TRPO problem as follows:
\begin{align}
    \max_{\theta}\frac{1}{T}\sum_{t=0}^{T}\bigg[\frac{\pi_{\theta}(a_t|s_t)}{\pi_{\theta_{old}}(a_t|s_t)} A_t\bigg] \nonumber \\
    \text{subject to } D_{KL}^{\theta_{old}}(\pi_{\theta_{old}},\pi_{\theta}) \leq \eta  \label{eq19}
\end{align}
where $A_n$ is the advantage function calculated by Generalized Advantage Estimation (GAE) ~\cite{schulman2015high} below:
\begin{align}
     & A_t = \sum_{l=0}^\infty (\gamma \lambda_g)^l \delta_{t+l}^V \nonumber \\
     & \text{  where  } \delta_{t+l}^V = -V(s_t) + \sum_{l=0}^\infty \gamma^l r_{t+l} \label{eq20}
\end{align}
where the reward $r_{t+l}$ is the $l$-step's test reward at timestamp $t$. The reward $r_{t+l}$ have two components which are reward returned by environment and the bonus reward calculated by Discriminator $D$ by using $\log(D(s_t,a_t))$.
Considering the massive computation load of updating the TRPO via optimizing Eq.(\ref{eq19}),
we use Proximal Policy Optimization (PPO)~\cite{schulman2017proximal} with the objective function below, to update the policy:
\begin{align}
    \mathbb{E}_t\Big[\min\Big(\frac{\pi_{\theta}(a_t|s_t)}{\pi_{\theta_{old}}(a_t|s_t)}A_t,\text{clip}\Big(\frac{\pi_{\theta}(a_t|s_t)}{\pi_{\theta_{old}}(a_t|s_t)},1-\epsilon,1+\epsilon\Big)A_t\Big)\Big] \label{eq21}
\end{align}
where $\epsilon$ is the clipping parameter, which represents the maximum percentage of change that can be updated at a time.

The training procedure involves two components: the discriminator and the actor-critic network.
The training algorithm is illustrated in Algorithm ~\ref{alg:d}.
Specifically, for the discriminator,
We use Adma as the optimizer to find the gradient for Eq.(\ref{eq17}) for weight $w$:
\begin{align}
    \mathbb{E}_{\pi}[\nabla_w\log(D_w(s,a))] + \mathbb{E}_{\pi_E}[\nabla_w\log(1-D_w(s,a))] \label{eq22}
\end{align}

\begin{algorithm}[ht]
\SetAlgoLined
 \SetKwInOut{Input}{input}
 \Input{Expert Policy $\pi_E$, current state $s$}
 Sampling expert trajectories $(s,a_E)\sim\pi_E$\; 
 Initialize discriminator parameter $w_0$ \;
 Initialize policy parameter $\theta_0$ \;
 Initialize clipping parameter $\epsilon$\;
 \For{$i= 0,1,\cdots$}{
  Sampling  trajectories $(s,a)\sim\pi_{\theta_i}$\; 
  Update the parameter $w_i$ by gradient on Eq.(\ref{eq22}) \; 
   \For{$k= 0,1,\cdots$}{
    Get the trajectories $(s,a)$ on policy $\pi_\theta = \pi(\theta_k)$ \;
    Estimate advantage $A_t$ using Eq.(\ref{eq20})\;
    \nosemic Compute the Policy Update \;
    \nonl\begin{equation*}
    \theta_{k+1} = \argmax_{\theta} \text{Eq.}(\ref{eq21})\;
    \end{equation*}
    \nonl By taking $K$ step of minibatch SGD (via Adma)\;
   }
   $\theta_i \leftarrow \theta_K$\;
 }
 \caption{Training algorithm for our model}
 \label{alg:d}
\end{algorithm}

\section{Experiments}
We report experimental evaluation of our model on a real-world online retail environment, VirtualTB~\cite{shi2019virtual}, on OpenAI gym\footnote{https://gym.openai.com}.
\begin{figure*}[ht]
    \begin{subfigure}{0.34\linewidth}
        \includegraphics[width=\linewidth]{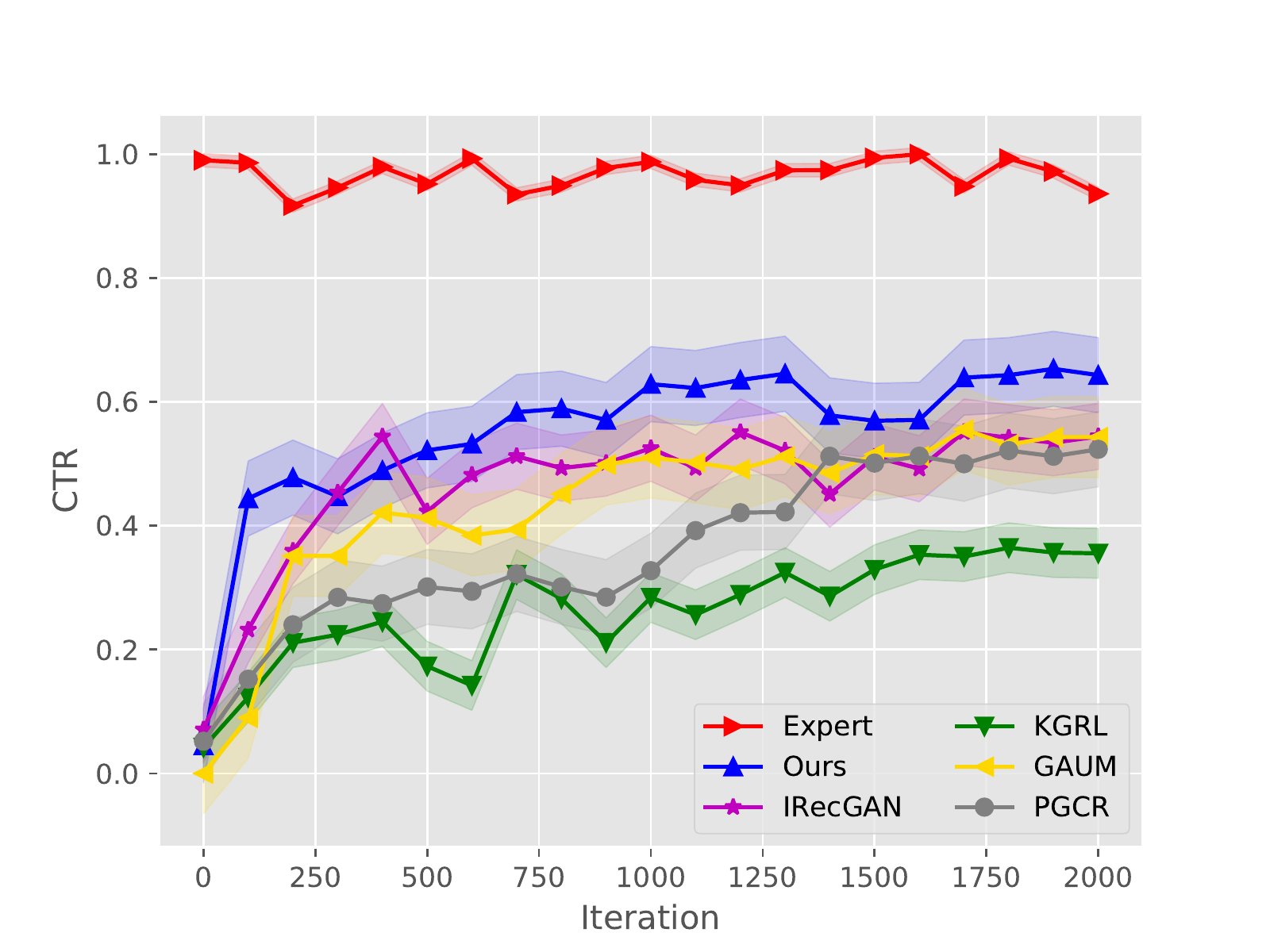}
        \caption{}
    \end{subfigure}
    \begin{subfigure}{0.34\linewidth}
        \includegraphics[width=\linewidth]{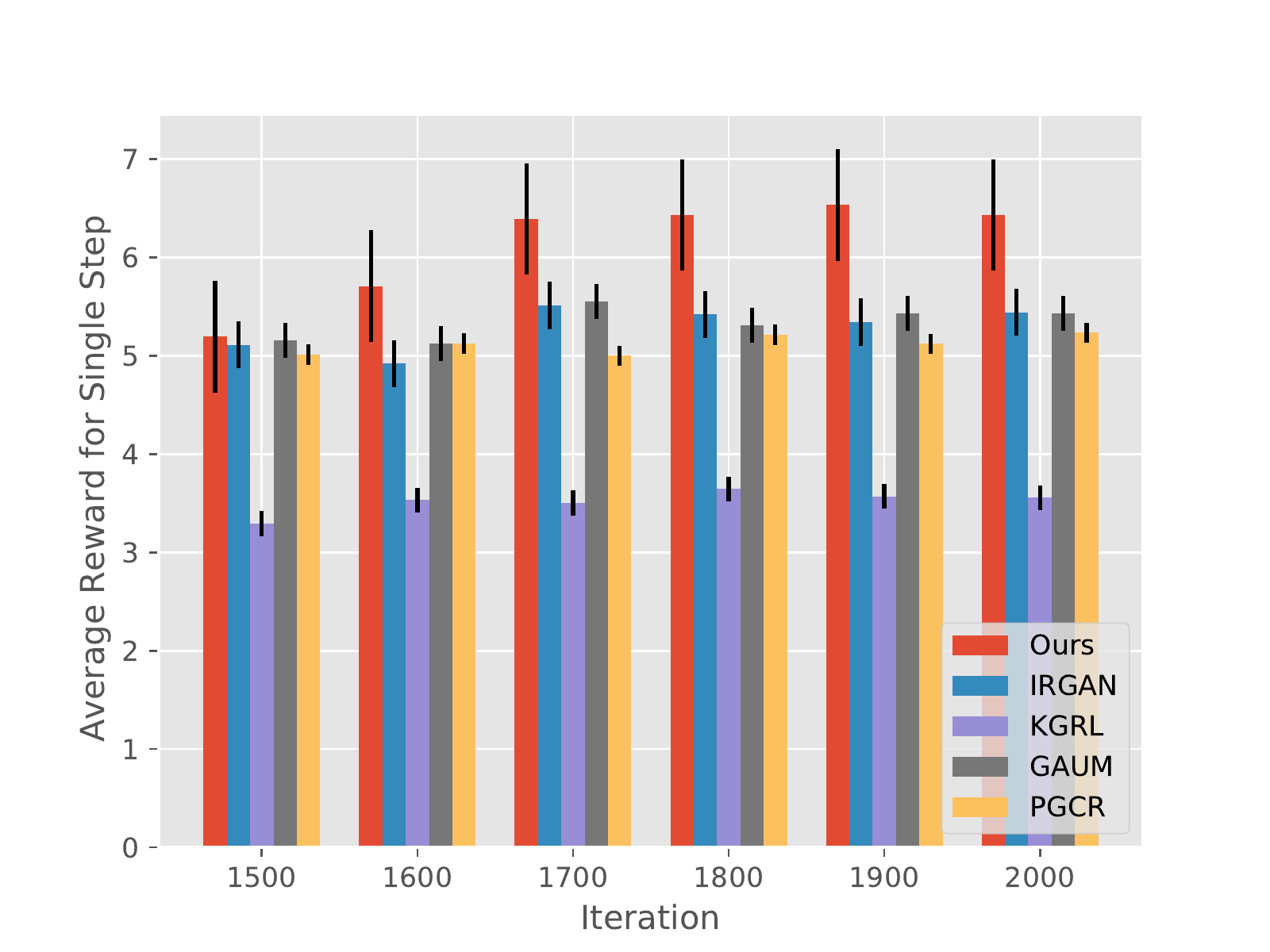}
        \caption{}
    \end{subfigure}
    \begin{subfigure}{0.34\linewidth}
        \includegraphics[width=\linewidth]{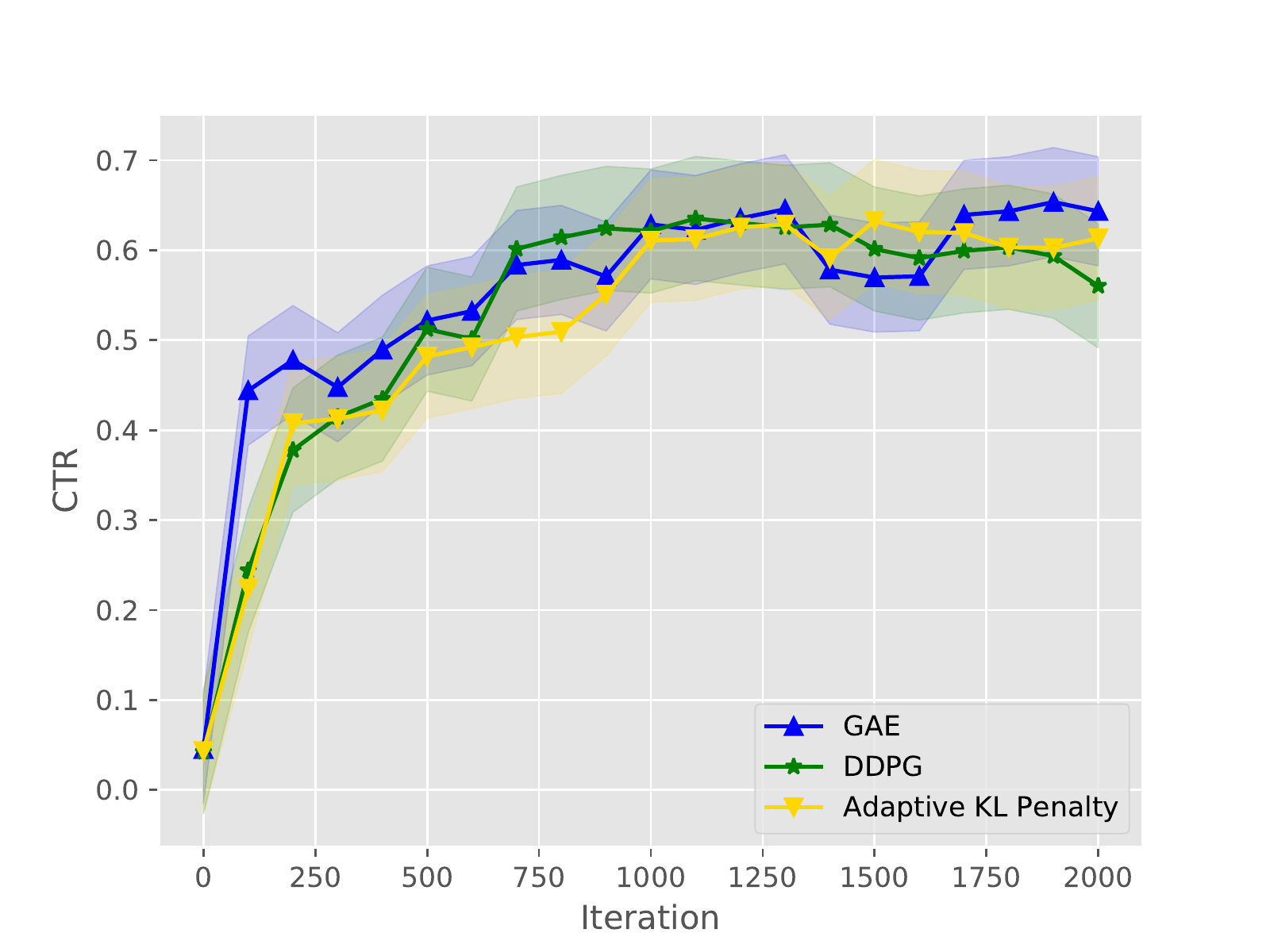}
        \caption{}
    \end{subfigure}
    \caption{Experimental results where (a) is CTR with 95\% confidence interval, and (b) is the average reward received each step after 1500 iterations with 95\% confidence interval. For comparison, we have added the performance from expert policy to (a). Results reported in (c) is for ablation study.}
    \label{fig:result}
\end{figure*}

\subsection{Virtual TaoBao}
VirtualTB is a dynamic environment to test the feasibility of the recommendation agent. It enables a customized agent to interact with it and achieve the corresponding rewards.
On VirtualTB, each customer has 11 static attributes as the demographic information and are encoded into a 88-dimensional space with binary values.
The customers have multiple dynamic interests that are encoded into a 3-dimensional space and may change over interaction process.
Each item has several attributes, e.g,. price and sales volume, and are encoded into a 27-dimensional space. 

\subsection{Baseline methods}
\begin{itemize}
    \item IRecGAN~\cite{bai2019model}: An online recommendation method that employs reinforcement learning and GAN.
    \item PGCR~\cite{pan2019policy}: A policy Gradient based method for contextual recommendation.
    \item GAUM~\cite{chen2019generative}: A deep Q-learning based method that employs GAN and cascade Q-learning  for recommendation.
    \item KGRL~\cite{chen2020knowledge}: Actor-Critic based method for interactive recommendation, a variant of online recommendation.
\end{itemize}
Note that GAUM and PGCR are not designed for online recommendation, and KGRL requires knowledge graph as side information, which is unavailable to the gym environment. Hence, we only keep the network structure and put those network into the VirtualTB platform for testing.

\subsection{Evaluation Metric and Experimental Environment}
The experiments are conducted in the OpenAI gym environment where the reward can be readily obtained for each episode. Since each episode may have different number of steps, it leads to the difficulty in determining when users will end the session. For this reason, we choose click-through-rate to represent the performance which is defined as:
\begin{align}
    CTR = \frac{r_{episode}}{10*N_{step}}
\end{align}
where $10$ means that user is interested in all 10 items which are recommended in a single page, $r_{episode}$ is the reward received per episode and $N_{step}$ is number of step included in one episode.

The model is implemented by using PyTorch~\cite{paszke2019pytorch} and the experiments are carried out on a server which consists of two 12-core/ 24-thread Intel (R) Xeon (R) CPU E5-2697 v2 CPUs, 6 NVIDIA TITAN X Pascal GPUs, 2 NVIDIA TITAN RTX, with a total 768 GiB memory.
\subsection{Expert Policy Acquisition}
In this part, we introduce the strategy on acquiring the expert policy for VirtualTB. There is no official expert policy in VirtualTB. Obviously, it is unrealistic to manually create the expert policy from this virtual environment where the source data is not available. Hence, we follow the similar strategy as in ~\cite{gao2018reinforcement} to generate a set of expert policy from a pre-trained expert policy network. 
We design an actor-critic network $\mathcal{M}$ with the same actor and critic network structure as our model, but without advantage. The critic network from $\mathcal{M}$ is used to calculate the $Q$-value by adopting deep Q-learning. We adopts the Deep Deterministic Policy Gradients (DDPG)~\cite{lillicrap2015continuous} to train $\mathcal{M}$. 

\begin{table*}[ht]
\caption{CTR for Different Parameter Settings for GAE and PPO with 95\% Confidence Interval}
\centering
\begin{tabular}{c|c|c|c|c|c|c|c}
    \hline
    \multicolumn{2}{c|}{\multirow{2}{*}{}} & \multicolumn{6}{c}{GAE: $\lambda_g$}      \\ \cline{3-8} 
    \multicolumn{2}{c|}{}                 & 0.94 & 0.95 & 0.96 & 0.97 & 0.98 & 0.99 \\ \hline
    \multirow{6}{*}{PPO: $\epsilon$} 
    & 0.05& 0.630 $\pm$ 0.063& 0.632 $\pm$ 0.064& 0.633 $\pm$ 0.062 & 0.630 $\pm$ 0.059 & 0.626 $\pm$ 0.060 & 0.629 $\pm$ 0.059 \\
                &  0.10& 0.632 $\pm$ 0.062 & 0.635 $\pm$ 0.060& 0.636 $\pm$ 0.061 & 0.636 $\pm$ 0.058 & 0.634 $\pm$ 0.061 & 0.633 $\pm$ 0.060 \\ 
                &  0.15& 0.633 $\pm$ 0.060 & 0.635 $\pm$ 0.061& 0.639 $\pm$ 0.061 & 0.640 $\pm$ 0.057 & 0.639 $\pm$ 0.059 & 0.638 $\pm$ 0.061 \\ 
                &  0.20& 0.634 $\pm$ 0.060 & 0.636 $\pm$ 0.060& 0.641 $\pm$ 0.063 & \textbf{0.643 $\pm$ 0.061} & 0.643 $\pm$ 0.063 & 0.641 $\pm$ 0.058 \\ 
                &  0.25& 0.631 $\pm$ 0.061 & 0.635 $\pm$ 0.059& 0.636 $\pm$ 0.060 & 0.637 $\pm$ 0.060 & 0.636 $\pm$ 0.061 & 0.634 $\pm$ 0.059 \\ 
                & 0.30 & 0.630 $\pm$ 0.059 & 0.631 $\pm$ 0.061 & 0.632 $\pm$ 0.060 & 0.630 $\pm$ 0.059 & 0.630 $\pm$ 0.058 & 0.629 $\pm$ 0.050\\ \hline
\end{tabular}
\label{tab:result}
\end{table*}
\subsection{Hyper Parameters Setting}
For the policy network $\mathcal{M}$, we set the DDPG parameters as: $\gamma = 0.95, \tau=0.001$, size of hidden layer is $128$, the size of reply buffer is $1,000$ and the number of episode is set to $200,000$. For Ornstein-Uhlenbeck Noise, scale is $0.1$, $\mu=0, \theta=0.15, \sigma=0.2$. For our approach, number of episode is set to $100,000$, hidden size of the advantage actor-critic network is $256$, hidden size for discriminator is $128$, learning rate is $0.003$, factor $\lambda$ is $10^{-3}$, mini batch size is $5$ and the epoch of PPO is $4$. For the generalized advantage estimation, we set the discount factor $\gamma$ to $0.995$, $\lambda_g = 0.97$ and $\epsilon = 0.2$. For fair comparison, all those baseline methods are training under the same condition. For easy recognition, we set one iteration as 100 episodes.

\subsection{Results}
Fully results are reported in Fig~{\ref{fig:result}}. Our approach generally outperforms all four state-of-the-art methods. Specifically, our method gets a best result over all those baseline methods after $500$ iterations. It demonstrates the feasibility of the proposed approach.A possible reason for the poor performance of KGRL is that KGRL maintains a local knowledge graph inside the model and actively interacts with the environment. Because the experiments are conducted on an online platform which does not provide the side information for KGRL to generate its knowledge graph. Hence, KGRL performs poorer than other baseline methods. 
\subsection{Impact of Key Parameters}
We are interested in how the control parameter $\lambda$ on GAE and the clipping parameter $\epsilon$ on PPO affect the performance.  These two key parameters  significantly affect the generalized advantage estimation and proximal policy optimization process. The $\lambda$ is used to make a compromise between bias and variance which normally is selected from $[0,95,1)$ with step 0.01. The clipping parameter $\epsilon$ is used to determine the number of percentage need to be clipped, normally smaller than $0.3$ to control the optimization step size. The results are reported in Table~\ref{tab:result}. For fairness, we report CTR at iteration 2000. Observe that when $\lambda=0.97$ and $\epsilon=0.2$, the model achieves the best result which is $0.643 \pm 0.061$. These two values are also used as our default setting. More details about the model $\mathcal{M}$ can be found in the {\em 
Supplementary Materials}.
\subsection{Ablation Study}
In this part, we investigate the effect of the GAE. We use two different optimization strategies to optimize the proposed model which are DDPG and Adaptive KL Penalty Coefficient. The Adaptive KL Penalty Coefficient is the simplified version of the PPO which can be defined as:
\begin{align}
    L(\theta) = \mathbb{E}_{t}\Big[ \frac{\pi_{\theta}(a_t|s_t)}{\pi_{\theta_{old}}(a_t|s_t)}A_t - \beta \text{KL}[\pi_{\theta_{old}}(\cdot|s_t),\pi_{\theta}(\cdot|s_t)]\Big]
\end{align}
The update rule for $\beta$ is:
\begin{align*}
     \begin{cases} 
      \beta \leftarrow \beta/b & d < d_{target} * a  \\
      \beta \leftarrow \beta*b & d >= d_{target} * a \\
   \end{cases}\\
   \text{where } d = \mathbb{E}_{t}[\text{KL}[\pi_{\theta_{old}}(\cdot|s_t),\pi_{\theta}(\cdot|s_t)]]
\end{align*}
The parameter $a=1.5$, $b=2$ and $d_{target} = 0.01$ are determined by experiments where the selection process are reported in {\em 
Supplementary Materials} due to the space limit. The result of the ablation study can be found on Fig~\ref{fig:result} (c).

\subsection{Discussion}
This study provides a new approach for reinforcement learning based online recommendation, without the need of defining reward function. In this way, our work is feasible to be applied in various real-world recommendation scenarios, where the reward function is hard to manually define or highly domain-dependent. The proposed method offers a fundamental support for inverse reinforcement learning based recommendation system. By providing a few user behaviors, the proposed method can extract an adaptive unknown reward function so as to automatically find out the optimal strategies to generating recommendations best fitting users' interest. Our empirical evaluation testify its competitive performance against reinforcement learned based existing state-of-the-art methods.
Our model has implication and potentially accelerates the progress in applying reinforcement learning in practice where a complex environment exists.

\subsection{Related Work}
We briefly review previous studies related to deep reinforcement learning (DRL)-based recommendation. All those methods are MDP-based or partial observable MDP-based (POMDP). POMDP based methods can be further divided into three categories: value function estimation~\cite{hauskrecht1997incremental}, policy optimization~
\cite{poupart2005vdcbpi}, and stochastic sampling~\cite{kearns2002sparse}. Due to the high computational and representational complexity of the POMDP based methods, MDP-based methods are relatively more popular in academia.

MDP-based DRL methods for recommendation can be concluded in three different approaches: deep Q-learning based, policy gradient based and Actor-Critic based methods. \cite{zheng2018drn} adopts the deep Q-learning into the news recommendation by using user's historical record as the state. \cite{zou2020pseudo} improves the structure of Deep Q-learning to achieve more robust results.
\cite{pan2019policy} applies the policy gradient to learn the optimal recommendation strategies. \cite{wang2020kerl} introduces the knowledge graph into the policy gradient for sequential recommendation. However, Q-learning may get stuck because of the max operation, and policy gradient requires a large scale data to boost the converge speed and it will only update once per episode. Hence, 
Actor-Critic uses the Q-value to conduct the policy gradient per step instead of episode. \cite{zhao2017deep} adopts the actor-critic methods to conduct the list-wise recommendation 
in a simulated environment. \cite{chen2020knowledge,zhao2020leveraging} utilize the knowledge graph as the side information embedded into the state-action space to increase model's capability on actor-critic network. In addition,~\cite{liu2020end} proposes to produce recommendations via learning a state-action embedding within the DRL framework. 

Furthermore, \cite{chen2019generative} integrates the generative adversarial network with reinforcement learning structure to generate user's attribute so that more side information would be available to boost the reinforcement learning based recommendation system's performance. \cite{shang2019environment} proposes a multi-agent based DRL method for environment reconstruction which take the environmental co-founders into account.

\section{Conclusion and Future Work}
In this paper, we propose a new approach InvRec for online recommendation. Our proposed approach are designed to overcome the drawback due to the inaccurate reward function. The proposed model is built upon advantage actor-critic network with the generate adversarial imitation learning. We evaluate our method on the online platform VirtualTB and our model achieves a good performance. We also compared our method with a few state-of-the-art methods in three different categories: Deep Q-Learning based, policy gradient based and actor-critic network based methods. The results demonstrate that the proposed approach's feasibility and superior performance. 

This study provides a good initial attempt about the application of deep inverse reinforcement learning on online recommendation system. However, there are remains a few shortcomings which are not addressed in this paper such as the sample inefficiency problem for the imitation learning ~\cite{kostrikov2018discriminatoractorcritic}. Low sample inefficiency will lead to longer training time and the need of a larger dataset. Random sampling also would affect the performance. The possible solutions would be using the off-policy methods instead of on-policy, finding an optimal sampling strategy such that agent will get the same expert trajectories when facing the state which comes up before.

\bibliography{sample-base}
\end{document}